\documentclass{elsart}
\journal{New Astronomy}

\usepackage{natbib}

\usepackage{graphicx}
\usepackage{epsfig}

\usepackage{amssymb}

\begin{document}

\begin{frontmatter}



 \title{First Light and Reionization: A Conference Summary}


\author{E. J. Barton, J. S. Bullock, A. Cooray, M. Kaplinghat}
\address{Center for Cosmology, University of California, Irvine,
4129 Frederick Reines Hall, Irvine, CA 92697-4575}

\begin{abstract}
The search for the first illuminated astronomical sources in the
universe is at the edge of the cosmic frontier.  Promising techniques
for discovering the first objects and their effects span the
electromagnetic spectrum and include gravitational waves.  We
summarize a workshop on discovering and understanding these sources
which was held in May 2005 through the Center for Cosmology at the
University of California, Irvine.

\end{abstract}

\begin{keyword}
cosmology: theory \sep cosmology: observations \sep intergalactic medium 
\sep galaxies: formation \sep galaxies: evolution
\end{keyword}

\end{frontmatter}

\section{Introduction}
\label{intro}

In May 2005, 55 international researchers convened in Irvine to
discuss the search for first-light objects.  Blending radio, infrared,
optical, x-ray, and gravitational wave observers with theorists
working on all aspects of structure and galaxy formation, the breadth
of the conference and the variety of observational techniques being
employed generated optimism for the prospect of understanding first
light in the coming decades.

The current observational evidence paints a challenging (yet partially
complete) picture for our theoretical colleagues.  One hand we have
the striking {\it Wilkinson Microwave Anisotropy Probe (WMAP)} result,
suggesting a relatively high Thompson optical depth \citep[$\tau =
0.17 \pm 0.04$;][]{Kogut2003} and an early epoch of reionization, $z
\sim 15$.  On the other, we have spectra from Sloan Digital Sky Survey
(SDSS) quasars giving spectacular evidence for at least a partially
neutral IGM at a much later epoch ($z\sim 6.5$).  If true, the early
onset for reionization suggested by WMAP seems to
require\footnote{unless LCDM itself is flawed, e.g.  Cen, this volume}
a Pop III IMF weighted toward massive stars
\cite{Cen2005,Tumlinson2005}.  Reconciling the two observations with
each other demands an extended reionization, driven perhaps by the
transition from Pop III to Pop II star formation itself
\cite{Cen2005,Venkatesan2005,Schneider2005}. Indeed, these empirical
indications follow qualitatively in line with expectations from those
working on the theory of first star formation
\cite[e.g.][]{Yoshida2005}.

Direct observations of galaxies present a similarly intriguing
picture.  The presence of massive, old galaxies at $z\sim 6$
\cite[e.g.,][]{Yan2005a,Yan2005b} may pose a serious challenge to LCDM
theory, but more likely provides insight into the nature and rate of
star formation in the rarest objects \cite{Dave2005,Nagamine2005}.
Indeed, evidence for cosmic ``downsizing'' --- that star formation in
massive galaxies peaked at an earlier time than star formation in
low-mass galaxies --- suggests that fundamental understanding of the
bulk of the first stars may rely on our understanding of star
formation in the most massive systems
\cite[e.g.,][]{Abraham2004,Papovich2005}.  Also puzzling, in light of
evidence for early reionization, are indications for a decline in
global star formation from $z\sim 3$ to $z\sim 6$ \cite{Bunker2004,
Bunker2005, Bouwens2005b,Bouwens2005c}.

Cosmological hydrodynamic simulations probe, in some detail, the
nature of the global star formation history and of high redshift
galaxies themselves \cite{Dave2005,Nagamine2005,Yoshida2005}.  While
predictions for stellar masses and even photometric properties are
relatively robust given the model assumptions, more detailed
predictions for many properties of galaxies at end of the dark ages
epoch remain elusive.  The inability to predict the exact nature and
timing of reionization and the Lyman $\alpha$ fluxes of high redshift
galaxies arises primarily from uncertainties in the escape fractions
of ionizing photons, the stellar IMF at different epochs, and the
nature of feedback in low-mass galaxies at high redshift.

Because of the uncertainties in model predictions, progress relies on
advancing our observational knowledge of the high-redshift universe.
It will come only through deeper, more exhaustive searches for early
star-forming objects \cite[e.g.,][]{Windhorst2005, Willis2005a,
Willis2005b, Barton2004,Stark2005} and their signatures
\cite[e.g.,][]{Kashlinsky04, Carilli05}, and a continued understanding
of the role of gaseous structures and gas infall into galaxies
\cite[e.g.,][]{Cooke2005}.  The current generation of astronomical
facilities is poised to address many of the important questions.  In
addition, with the advent of the next generation of ground- and
space-based facilities, including {\it JWST}, TMT, GMT, and SPICA,
photometric techniques and Lyman $\alpha$ searches will bring us to a
point where we are obtaining information on large numbers of galaxies
inhabiting the earliest epoch of star formation in the universe.
Near-field searches for fossil relics of the first stars will provide
local probes of these early processes.  Improved measurements of
fluctuations in the infrared background will reveal early stars, and
21 cm measurements of neutral hydrogen in the universe will unveil the
process of reionization directly.  Thus, although many mysteries
surround the cosmic ``dark ages'' at present, the future of the field
is very bright.

\section{Observations: Evidence for Early Galaxies and Structure}
\label{galobs}

Although many questions remain, existing observations offer a wide variety 
of constraints on the processes of early galaxy formation and
cosmic reionization.

\subsection{The Cosmic Microwave Background and Reionization}
\label{cmb}

The lack of strong damping by electron scattering of the acoustic
peaks in the temperature anisotropy of the CMB radiation has shown
that the universe was neutral approximately between the redshifts $30
< z < 10^3$. The classic way to test for the presence of neutral
hydrogen is to look for spectra of distant background objects such as
quasars that have troughs consistent with no flux at a substantial
stretch of wavelength shortward of
$(1+z)\lambda_\alpha=8850$\AA. However, because of the large opacities
of the hydrogen Lyman series, it is difficult to push this method to
directly infer the presence of neutral fractions of order unity.

The cosmic microwave background (CMB), on the other hand, is sensitive
to the ionized electron fraction through Thomson scattering of photons
\cite{bond84,polnarev85,zaldarriaga95}. This allows the CMB to
estimate the optical depth to Thomson scattering \cite{zaldarriaga96}
as well as its variation with time \cite{kaplinghat02}.  Thus, results
from the CMB have played a major role in the study of reionization in
recent years.

In his talk, Brian Keating explained how the large angle polarization
correlations may be used to obtain the optical depth to Thomson
scattering. Thomson scattering of anisotropic radiation results in a
net linear polarization produced locally \cite{Keating2005}. The resulting
polarization anisotropy appears at large angles because the horizon at
reionization is much larger than that at recombination. Keating
explained how the polarization anisotropy may be identified as B-mode
(gradient-free) or E-mode (curl-free). Scalar fluctuations only give
rise to E-mode polarization in linear theory. Keating emphasized that
the CMB could answer questions such as: what is the optical depth to
the last scattering surface, was reionization gradual or
instantaneous, and how inhomogeneous was it?

The optical depth measurement is also important for other fundamental
reasons. It is key to determining the amplitude of the primordial
scalar fluctuations \cite{zaldarriaga97} and may play a decisive role in
constraining or measuring the amplitude of the tensor (gravity wave)
perturbations \cite{kaplinghat03}.

At the other end of the spectrum are the signatures of patchy
(inhomogeneous) reionization on small angular scales, discussed by
McQuinn \cite{mcquinn05a,mcquinn05b}. The patchy reionization signal
results from correlations of the perturbations of the ionization
fraction and the large-scale bulk flow. This is comparable to or
larger than the kinetic SZ (kSZ) signal resulting from non-linear
baryonic density perturbations caught up in the bulk flow. Modeling
correlations of ionized bubbles {\it ab initio} is hard. McQuinn
pointed out that our current understanding of reionization suggests
that overdense regions collapse first and that the bubbles tend to
be large, $\sim 10 {\rm Mpc}$. McQuinn used a model \cite{furlanetto04}
inspired by the Press-Schechter framework to predict the secondary CMB
anisotropy signature from the patchy reionization effect. McQuinn
finds that around $\ell \sim 1000$, the patchy reionization signal is
comparable to the kSZ signal. He emphasized that it may be possible to
discriminate between different dominant sources of reionization (black
holes vs pop III stars) using the patchy reionization signal. He also
emphasized kSZ will be a significant contaminant for cosmological
parameter estimation from primary and lensed CMB \cite{santos03}.

Wright discussed the thermal SZ signal on these small scales. He
pointed out that if the small scale excess temperature correlations
seen by CBI and ACBAR are due to thermal SZ, then a $\sigma_8$ on the
high side will be required. He pointed out that in this case, a high
optical depth $\tau$ will be favored since $\sigma_8 \exp(-\tau)$ is
well determined from the CMB primary anisotropy. This would be
consistent with the WMAP large angle determination of the optical
depth.

In order to identify the small-scale ``excess'' as thermal SZ, one
must observe it in multi-frequency bands. Experiments like ACT and SPT
will do just that in the near future, and try to look for the thermal
SZ frequency distortion. Wright also discussed other frequency
distortions in the CMB frequency spectrum due to hot electrons. He
showed that the constraints on the y-parameter from FIRAS limits the
temperature of the electrons to less than about 50 eV.

DASIPOL first measured the CMB polarization angular
power spectrum. DASIPOL results are consistent with the TE
cross-correlation spectrum obtained by WMAP. While DASIPOL
could not see the large angle correlations, WMAP could and found
the TE cross-correlation higher than expected which suggested that the
universe may have been reionized earlier than thought. Wright discussed
these points in more detail. He pointed out that with all-sky
polarization data we were entering a new regime where foregrounds
(like polarized radiation from synchrotron and dust emission)
will no longer be sub-dominant. Wright pointed out that the E mode
auto-correlation (which WMAP is attempting to measure now) is much
smaller that the TT or TE signal at about 0.3 $\mu{\rm K}$. This will
be a difficult measurement given the galactic foregrounds
whose effects are hard to quantify.

\subsection{Quasars, gamma ray bursts, and the transparency of the IGM}

In a now-famous study of high-redshift quasars selected from the Sloan
Digital Sky Survey (SDSS), Gunn-Peterson troughs discovered at
$z \sim 6.2$ indicate that the IGM is at least partly
neutral at this redshift \cite{Becker2001, Fan2002}.  Both Xiaohui Fan
and George Djorgovski pointed out that high-z QSOs actually show
increase in IGM opacity starting at $z=5.5-5.7$ and a variation in IGM
transmittance along different lines of sight; the variation peaks just
above $z=5.5$ \cite{Djorgovski2005}.

The prospects for characterizing the IGM at high-$z$ are good for the
coming years.  Fan stated that as of May 2005, there are 9 $z > 6$
quasars from SDSS, with a total of 30 expected upon completion of
the survey.  Djorgovski indicated that the Palomar-Quest survey will
double the number of known high-redshift quasars in the next few years
\cite{Djorgovski2005}.  As pointed out by Djorgovski, 
Fan, and Andrei Mesinger, these quasars will offer more lines of sight for
IGM transmittance tests to probe the neutral regime
$\left<x_{HI}\right> \leq 0.01$ \cite{Mesinger2005}.

As emphasized by Don Lamb, Gamma ray bursts (GRB's) may provide a
promising near-term method of probing the IGM (and star formation
rate) to even higher redshifts.  Derek Fox described how with current
facilities, GRB afterglows can be observed to $z > 10$ in the
near-infrared with rapid follow-up.  Joe Jensen pointed out that in
semesters 2004B and 2005A, five teams had active programs to find
gamma-ray burst afterglows, primarily in the optical, taking advantage
of the telescope's rapid response capabilities \cite{Jensen2005}.
Follow-up work in the the infrared is beginning with Gemini.

\subsection{Distant Galaxies and the Lyman break technique}

The CMB appears to indicate an epoch of reionization as early as $z
\sim 20$.  The most distant astronomical object confirmed through both
broad-band imaging and spectroscopy lie a redshifts just under $z \sim
7$.  The cosmic dark ages ended at an epoch somewhere between these
two extremes.  As Xiaohui Fan indicated, the QSO density is down by
a factor of 40 from $z=2.5$ to $z=6$; in fact, the $z=6$ density of the
luminous quasars is only 1 per Gpc$^{-3}$.  Faint quasars are also not
abundant at these redshifts.  Fan pointed out that the deep SDSS strip
shows the high-redshift quasar LF is flat; at high $z$ there are fewer
low-luminosity quasars for each high-luminosity system.  Thus, quasars
did not reionize the universe \cite{Fan2001}.  Star-forming galaxies
are most likely the dominant sources of ionizing photons
\cite{Tinsley73}.  Thus, the search for the galaxies responsible for
reionization lies directly on the cosmic frontier.  We will not
understand the history of the early universe until we find and
understand the population of high-redshift galaxies that first bathed
the universe in ionizing radiation.

Many efforts are now underway to study the earliest galaxies.  The
``Lyman break'' technique, first successfully pioneered at $z \sim 3$
over a decade ago \cite{Steidel93}, finds high-redshift galaxies using
broad-band photometry sensitive to the Lyman continuum break.  Since
that time, the push to find ``Lyman break'' galaxies (LBGs) has
progressed to $z \sim 7-8$ and beyond \cite[e.g.,][]{Bouwens2005b}.
One key strength of the Lyman break technique is that ultraviolet
continuum emission is closely related to star formation rate.  This is
in contrast to Ly$\alpha$ line searches because resonance scattering
attenuates the Ly$\alpha$ line.  However, Lyman break identifications
and photometric redshifts can be ambiguous, UV continuum emission can
be absorbed by dust, and contamination from foreground sources is
possible.  In addition, like Ly$\alpha$ searches, LBG samples
typically include only rapidly star-forming galaxies.

The {\it Spitzer Space Telescope} is a key development allowing red
photometry that reveals accurate star formation histories for even
aging moderate-redshift galaxies.  Combining galaxy surveys including
the Spitzer/MIPS GTO surveys, the Great Observatories Origins Deep
Survey \cite[GOODS; e.g.,][]{Giavalisco2004}, and COMBO-17, Casey
Papovich reported that most of the distant red galaxies (DRGs) at z =
1-3.5 are detected in 3-24 $\mu$m range \cite{Papovich2005}.
Resembling LIRGs and ULIRGs, many have vigorous star formation,
yielding a vast range of stellar masses and and star formation rates.
Both Papovich and Patrick McCarthy argued that the moderate-redshift
galaxies show evidence for the now-popular ``downsizing'' concept:
massive galaxies formed stars more actively in past, while the smaller
galaxies are more active now.  Reporting on results from the Gemini
Deep Deep Survey \citep[GDDS;][]{Abraham2004}, McCarthy also indicated
that there is no shortage of old, high-mass, evolved systems at $z
\sim 2$ (McCarthy et al. 2004).  Surveys show evident downsizing as
the star formation rate peaks for massive systems at $z > 2$, but for
low-mass systems at $z < 1$.  Papovich suggested that old stellar
populations in massive $2 < z < 3$ galaxies can account for 5\%-15\%
of LBGs at $z \sim 6$ for a standard IMF; a flat IMF would allow these
galaxies to explain 90\% of the z$\sim$6 ultraviolet luminosity
density\cite[see][]{Papovich2005}.

Jeff Cooke reported that the damped Lyman $\alpha$ system (DLA)-LBG
cross-correlation amplitude is consistent with the LBG
auto-correlation amplitude \cite{Cooke2005, Cooke2005b}.  Thus, DLAs
and LBGs may be the same population.  Because DLAs contain a large
fraction of the neutral hydrogen in the universe, their as-yet unknown
contributions to the ionizing flux in the universe may prove
important.  Cooke also reported on an ongoing survey to understand
DLAs at $3.5 \lesssim z \lesssim 5$.

Multiple researchers reported on the search for $z=6$ galaxies in the
Hubble ultra-deep field (HUDF) and similar HST surveys.  Andrew Bunker
reported on spectroscopic studies of I-band dropouts with confirmed
redshifts at $z \sim 6$, noting that the numbers of I-band dropouts
are consistent in the GOODS-N and GOODS-S fields.  The luminous
I-dropouts in the HUDF indicate that the measured star formation rate
at $z \sim 6$ is lower by factor of 6 from the $z \sim 3$ star
formation rate.  If the estimate is correct, the I-dropouts do not
emit enough ionizing photons to reionize the universe \cite{Bunker2004,
Bunker2005}.  The discrepancy could be solved by a steep faint-end
slope to the LF, a different stellar initial mass function (IMF) or
rapid star formation in galaxies at much higher redshifts.

Rychard Bouwens reported a somewhat higher ultraviolet luminosity at
$z \sim 6$, constructing a LF from 346 galaxies from the HUDF, the
UDF-Parallel, and the GOODS survey fields.  The LF extends 3
magnitudes fainter than L$^{\star}$.  Bouwens reported significant
evolution in L$^{\star}$ -- a doubling from $z \sim 3$ to $z \sim 6$
\cite{Bouwens2005b}.  In contrast to the Bunker report and in closer
agreement with \cite{Giavalisco2004}, Bouwens et al. find a luminosity
density that is only a factor of 1.5 less than the luminosity density
at $z \sim 3$.  However, many of the low-luminosity sources are too
faint for spectroscopic confirmation \cite{Bouwens2005c}.

Sangeeta Malhotra reported on the GRism ACS program for Extragalactic
Science (GRAPES) survey, which has measured spectra for 15\% of the
galaxies in the HUDF including a galaxy at $z = 6.7$.  They detect a
clear overdensity in the HUDF at $z=5.7$ \cite{Malhotra2005}. The star
formation density there, $2-4 \times 10^{-2}$ M$_{\circ}$ Mpc$^{-3}$
yr$^{-1}$, is high enough for local reionization.

The long-wavelength constraints at $z \gtrsim 6$ provided by the {\it
Spitzer Space Telescope} have revolutionized the study of first light.
As found by Haojin Yan and other speakers (Richard Ellis, Bunker),
rapid star formation at high-redshift is strongly supported by the
clear existence of old, massive galaxies at $z \sim 6$.  Because of
the long-wavelength data, spectral energy distribution fitting at $z
\sim 6$ is possible; best-fit models sometimes have stellar masses as
high as a few $\times 10^{10}$ that started forming more than a few
100 Myr ago.  They have metallicities that are consistent with solar
but with negligible dust.  Yan pointed out that some galaxies are too
blue to fit the models, which may be explained by the presence of $>
100$ M$_{\circ}$ stars \cite{Yan2005a, Yan2005b}.  In any case, there
is little doubt that star formation began before $z \sim 7$ in these
systems.  Mobasher et al. \cite{Mobasher2005} describe a $J$-band
dropout from the HUDF detected by Spitzer for which the best-fit model
is a redshift of $z \sim 6.5$ and a stellar mass of $6 \times 10^{11}$
M$_{\circ}$ with little or no ongoing star formation that formed its
stars at $z > 9$.  The galaxy may have emitted enough ionizing photons
to reionize the area of the NICMOS HUDF, beginning the local
ionization process at $z \sim 15$ \cite{Panagia2005}.

Strengthening the case for an active epoch of galaxy formation at $z
\sim 6$, Christopher Conselice discussed the application of
concentration and asymmetry statistics to $z \sim 6$ galaxies in the
HUDF, showing that they occupy range of morphologies.  He reported
candidates for major mergers, with mergers deviating from the apparent
relationship between concentration and size.  He reported overall
major merger (pair) and asymmetry fractions at $z \sim 6$ of about 20
\%, which is similar to $2 < z < 3$.

Similar arguments about the advanced state of early ($z \sim 6$) star
formation rely on the discovery of enriched systems at these
redshifts.  Peng Oh and George Becker addressed the question of metal
enrichment at high redshift, announcing the Keck/HIRES detection of
O{\sc I} absorbers at $z > 6$, but noting that only isolated
absorbers, and not a forest, were detected.  Quasars provide
additional information about advanced metal enrichment at high
redshifts.  Xiaohui Fan pointed out that the large halos that host
these supermassive black holes must assemble and enrich to super-solar
metallicities by $z=6$.

The HUDF provides hints of constraints at $z \gtrsim 7$.  Bouwens et
al. \cite{Bouwens2004} find 5 $z$-band dropouts consistent with
$z=7-8$ galaxies; they assume one is a contaminant and 4 sources are
real, while 14 are expected if there is no evolution from $z=3.8$ to
$z = 7-8$.  Thus, the luminosity density is down by an estimated
factor of $\sim$5 in luminous galaxies \cite{Bouwens2004}.  Richard
Ellis described a planned HST/Spitzer survey that will target lensed
$z$-band and $J$-band dropouts to 1 magnitude deeper than the HUDF
around six cluster fields in $z$, $J$, and $H$ with 110 orbits; the
total area will equal almost 25\% of the area of the NICMOS UDF.  The
survey will probe the faint end of the luminosity function, to search
for evolution in the number counts and total luminosity density to $z
= 7-10$ \cite{Stark2005}.

Searches for $J$-band dropouts in the HUDF $(z \sim 10)$ yield 3
candidates, while 4.8 are expected if there is no evolution from $z
\sim 6$ \cite{Bouwens2005a}.  However, none of the $z \gtrsim 7$
candidates have been spectroscopically confirmed to date.  Although
Gemini/GNIRS failed to detect emission from a HUDF $z$-band dropout, a
spectrum by Ellis, Stark, Windhorst, \& Yan with a clear non-detection
of H-$\alpha$ argued for a $z = 6.5$ ID of the luminous galaxy.  Joseph
Jensen emphasized that the Pell\'{o} et al. $z=10$ discovery
\cite{Pello2004} has been refuted at Gemini by a non-detection in a
deep $H$-band image \cite{Bremer2004}.  Jensen also talked about other
Gemini LBG science including a follow-up study of a Postman et al.
ACS massive cluster survey and work on GNIRS spectroscopy of $z$-band
dropouts \cite{Jensen2005}.

\subsection{Distant galaxies and Lyman-$\alpha$ emission-line searches}

Many of the most distant LBG sources are too faint for identifying
follow-up spectra with present-day technology.  Line emission provides
a distinct spectral feature that can be used to achieve a very solid
redshift for a distant galaxy.  Supporting this technique are
observations that indicate Lyman $\alpha$ emission is common in
high-redshift galaxies.  For example, reporting on results from the
GLARE project, both McCarthy and Bunker emphasized that 1/3 of the $z
\sim 6$ I-band dropouts have strong Lyman $\alpha$ emission
\cite{Stanway2004}.  Lyman $\alpha$ emission is a narrow spectral
feature that can be sought between night-sky emission lines.  The
primary advantage of this technique is in its sensitivity:
ground-based searches between the strong OH lines in the sky are
sensitive to even moderately star-forming galaxies if Lyman $\alpha$
is escaping.

Searches for Lyman $\alpha$ emission are ongoing at $z \sim 6$ and
beyond.  James Rhoads reported on the LALA survey for $z \sim 6.5$
Lyman $\alpha$ emitters, which uses the NOAO 4-meter telescopes and
stacks $\sim$24 hours of data.  At its highest redshift, $z \sim 6.6$,
LALA has surveyed a volume of $1.5 \times 10^{5}$~Mpc$^3$ to a depth
of $2 \times 10^{-17}$ erg s$^{-1}$ cm$^{-2}$ \cite{Rhoads2004}.
The studies report high Ly$\alpha$ equivalent widths that may require
a top-heavy IMF, low metallicity, or power from an active galactic
nucleus to explain; stacked x-ray images suggest that AGN play a role
in $\lesssim$20 \% of the sources.  Because the $z=5.7$ and $z=6.6$
Lyman $\alpha$ luminosity functions are consistent with one another,
Rhoads reports that there is no evidence that the universe is
significantly neutral ($\sim$30\%) at $z=6.6$.

Crystal Martin reported on a complementary multislit Lyman $\alpha$
survey to find $z=5.7$ emission with the Keck and Magellan to fainter
line fluxes than wide-field surveys such as LALA.  The Magellan survey
sensitivity is a few $\times 10^{10}$ erg s$^{-1}$ cm$^{-2}$; the
survey has produced 153 Lyman $\alpha$ candidates and 14 confirmed
candidates to date \cite{Martin2005}.

Moving beyond the optical regime and into the near infrared ($z
\gtrsim 7$) presents both sensitivity challenges and the possibility of
a neutral intergalactic medium (IGM) at these high redshifts.
Although neutral hydrogen scatters Lyman $\alpha$ very effectively,
theoretical arguments demonstrate that Lyman $\alpha$ photons can
redshift enough to escape if sources are placed inside locally ionized
bubbles, which might surround early strongly star-forming sources
\cite[e.g.,][]{Haiman2002, Santos2004}.  In fact, because Lyman $\alpha$
yields a direct probe of the local ionization state of the IGM,
wide-field surveys may ultimately help to map the topology of
reionization \cite{Furlanetto2005}.

Searches for Lyman $\alpha$ at $z \gtrsim 7$ are only beginning.
Daniel Stark described an ongoing search for Lyman $\alpha$ at $8 < z
< 10$.  With line sensitivities as low as $3 \times 10^{18}$ erg
s$^{-1}$ cm$^{-2}$ in some cases, the survey is sensitive to lensed
sources with star formation rates as low as $\sim$0.1 M$_{\circ}$
\cite{Stark2005}.  He reported 15 candidate high-redshift galaxies
that require confirmation.

Jon Willis reported on the highest-volume $z >7$ Lyman $\alpha$ survey
completed to date.  Using a narrow-band ($R=133$) 1.187 $\mu$m filter
in a $J$-band window between OH lines, Willis \& Courbin
\cite{Willis2005a} and Willis et al. \cite{Willis2005b} probe 340
h$^{-3}$ Mpc$^3$ at $z=8.76$ to a depth of $3.28 \times 10^{-18}$ erg
s$^{-1}$ cm$^{-2}$ in 35 hours on the VLT.  They detect 2
emission-line objects; both are ruled out as Lyman-$\alpha$ sources by
flux at bluer wavelengths.

John-David Smith described an ongoing project to search for Lyman
$\alpha$ with a narrow-band ($R=125$) filter he designed that lies
between OH lines at 1.122 $\mu$m ($z=8.227$) using Gemini/NIRI (see
Jensen 2005). The filter works extremely well, reducing the $J$-band
background by a factor of $\gtrsim 20$. With a target sensitivity of
just under $4 \times 10^{-18}$ erg s$^{-1}$ cm$^{-2}$, the probe
stands the best chance of success if there is efficient Population III
star formation at $z \sim 8$ \cite{Barton2004}.

None of the completed or ongoing surveys described at the conference
or in the literature have reported confirmed $z > 7$ candidates.
However, as these and other planned efforts advance and the parameter
space covered by blind and lensed searches expands, the likelihood of
finding a detection will increase dramatically in the coming years.

\subsection{The Extragalactic Background Light and IR Anisotropies}

Numerous arguments favor an excess contribution to the extragalactic
background light (EBL) between 1 $\mu$m and a few $\mu$m when
compared to the expectation based on galaxy counts and Milky Way
faint star counts \cite[see review in ][]{Hauser01}.  While these
measurements are likely to be affected by certain systematics and
issues related to the exact contribution from zodiacal light within
the Solar System, one explanation is that a contribution to the EBL
exists from high redshift sources in the form of an integrated
background made by Lyman-$\alpha$ emitting galaxies with the line
emission redshifted to the near-IR wavelengths observed today.

There are arguments both for and against such a population. The early
reionization observed by WMAP suggests the presence of UV-emitting
sources at redshifts greater than 10 or as early as a redshift of 20
in some scenarios for reionization. These sources could contribute to
the IR background (IRB) at wavelengths greater than 1 $\mu$m 
\cite[e.g.,][]{Cooray04b}.  Based on metallicity arguments at low
redshifts and background estimates at X-ray wavelengths, one can limit
the expected density of black holes or miniquasars at high redshifts.
These arguments \cite[e.g.,][]{Madau05} suggest that a large
density of primordial galaxies containing Population III stars cannot
be present at high redshifts and the allowed density based on
metallicity and reionization requirements is less than that required
to explain the missing IRB. On the other hand, while not explaining
the total missing IRB, even a small surface density of primordial
galaxies should leave an imprint between 1 and 5 $\mu$m 
\cite[see Figure~1 in ][]{Bock05}.

The study by Kashlinsky et al. \cite{Kashlinsky05} attempts to
understand the presence of a high redshift population through a deep
IR image taken with Spitzer IRAC as part of a GTO program at 3.6, 4.5,
5.8 and 8 $\mu$m.  With point sources removed down to 22.5 magnitudes
in the L-band, their analysis showed excess fluctuations beyond those
expected from known populations of galaxies with luminosities below
the point-source removal level. The fluctuations are clustered at
angular scales greater than 20 arcseconds though the measurement of
the IR anisotropy angular power spectrum is limited to the Spitzer
field of view of a few arcminutes.  These measurements probe some of
the deepest IR anisotropy measurements today and far exceed the depth
of previous observations with 2MASS from the ground
\cite{Kashlinsky04}, and experiments such as COBE DIRBE
\cite{Hauser98} and IRTS \cite{Matsumoto05} from space, or NITE with
rocket-borne IR imaging arrays \cite{Xu02}.

The excess fluctuations detected by Spitzer could be modeled based on
a population of high redshift sources following the models of Cooray
et al. \cite{Cooray04a} and are in good agreement with some of the
expectations for a fraction of the missing IRB.  Though Lyman-$\alpha$
emission is redshifted to near-IR the Spitzer imaging data does not
provide a strong probe of fluctuations related to Lyman-$\alpha$
emitting sources at redshifts between 8 and 15 because the low-end of
the wavelength studies is at 3.5 $\mu$m. In fact, the Spitzer result,
if due to a high-redshift population of galaxies, suggests the
existence of large anisotropies at lower wavelengths. The galaxies
could be present with luminosities fainter than the ones probed with
the Lyman-$\alpha$ luminosity function and could result in steepening
the faint-end slope, possibilities which remain to be established with
narrow-band filter observations \cite[details to appear in
][]{Cooray05}.

\subsection{Lessons from the high-redshift universe}

Synthesizing the high-redshift reionization required by WMAP with our
knowledge of galaxies in the high-redshift universe is still a job for
the future.  However, the data show clearly that $z \sim 6$ universe
was booming, containing sources with star formation and enrichment
histories that tell us star formation was active well before $z \sim
6$. In contrast to the basic premises of hierarchical formation
theory, however, evidence for ``downsizing'' appears to indicate that
the bigger galaxies formed most actively in the early epochs of the
universe.

There are some indications from high-redshift galaxy colors and Lyman
$\alpha$ equivalent widths that heavy stars may be required to explain
some sources at $z \gtrsim 5$ \cite[e.g.,][]{Yan2005a,Yan2005b,
Rhoads2004}.  However, there are still few direct constraints on the
high-redshift IMF, and it is possible that all the detections so far
are of Population II stars.

Other issues still surround the study of high-redshift galaxies.  For
example, the relationship between cosmic variance and the measured
luminosity functions of both high-redshift Lyman break galaxies and
Lyman $\alpha$ sources must be resolved.  How big does a survey need
to be to represent a fair sample of the universe?  For high-redshift
Lyman $\alpha$ sources, the answer to this question will depend not
only on the structures themselves, but on the ionization state of the
IGM. As these questions are resolved in the coming years, the issues
will naturally gravitate to more detailed questions regarding the
nature of the exact progenitors of present-day galaxies, their
detailed internal formation histories, and the role of the extremely
early history --- the role of reionization and possibly of Population
III enrichment --- in ultimately deciding the structure and
composition of a present-day galaxy.

\section{The Theory of Reionization and Early Structure Formation}
\label{galform}

If the WMAP result of a high Thompson optical depth holds true then
the standard LCDM paradigm is presented with at least two difficult
puzzles.  The first is simply achieving this level of optical depth at
all within the LCDM framework, and the second is explaining the
extended span of reionization $z \simeq 20-6$ required to match the
SDSS quasar results.  For example, Cen emphasized that the high WMAP
optical depth requires a very high star formation efficiency {\it and}
a high escape fraction of ionizing photons {\it and} an initial IMF
biased towards high-mass stars.  These empirical constraints along
with theoretical expectations \cite{Yoshida2005} point to a need for
an early IMF that favors the formation of high-mass stars and
efficient production of ionizing photons.  Interestingly, ultra
massive stars are not necessarily required to explain the observations
and a more moderate shift in the production of massive stars may be
the favored scenario for Pop III formation \cite{Tumlinson2005}.
Somewhat less massive early stars of this kind are perhaps favored by
the simulations of Yoshida, which include HD line cooling.

Schneider and Venkatesasn discussed
a scenario for transitioning from Pop III star formation to
a more typical Pop II IMF.
Specifically,  once a local  reservoir of gas  is enriched beyond a
specific critical metallicity ($Z_{cr}  \simeq 10^{-4}$ or  so) then  the IMF
switches from a top-heavy to a standard Pop II IMF.  Interestingly, in
this picture, the metal  ejecta  associated with the first   relatively
massive Pop III  stars provide chemical   feedback by enriching  the
local medium to a point above $Z_{cr}$.   This produces a reduction in
the  typical mass  of  star formed, and   thus naturally leads to more
inefficient conversion of baryons to ionizing photons.

It is possible that this kind of chemical feedback plays the important
role in extending the epoch of reionization.  Indeed maintaining a
partially ionized universe over a range as large as $z=20-6$ almost
certainly requires some level of astrophysical feedback.  ~\footnote{
Specifically, since structure formation only increases with time, this
suggests an astrophysical feedback source which regulates the
ionization fraction for an extended redshift range.}  Two avenues
naturally present themselves: 1) regulate star formation itself or 2)
regulate the production of ionizing photons.\footnote{ Another
possibility might involve stellar blow-out regulating the clumping of
gas and therefore the escape fraction of ionizing photons, see,
e.g. Yoshida, these proceedings} For example, the ionizing background
itself may regulate the formation of stars via suppressing the
formation of stars in low-mass galaxies.  The relevant ionizing field
may be due to local star formation or the ambient UV field itself, but
regardless the effect is to suppress the global star formation rate.
Alternatively, the metal ejecta associated with the first relatively
massive Pop III could provide the ``chemical feedback" by enriching
the local medium to a point above $Z_{cr}$ \cite{Schneider2005,
Venkatesan2005}.  This produces a reduction in the typical masses of
stars formed, thus reducing ionizing photons efficiency per stellar
mass.

The view in this meeting seemed to be weighted toward the role of
chemical feedback.  Dijkstra argued that photoionizing feedback may
not be enough to suppress star formation in mini-halos at the highest
redshifts because self-shielding is quite important at these
epochs~\cite{Dijkstra2005}.  Chemical feedback on the other hand seems
to arise naturally in any model where the $Z_{cr}$ paradigm is taken
seriously.  Depending on the degree of metal ejecta into the IGM
(which seems to be required at some level) the chemical feedback may
be a local phenomena, thus allowing the simultaneous existence of
massive PopIII stars and PopII stars at the same epoch (e.g.
Schneider, these proceedings).  This leads to the possibility that
metal free stars could be ``observed'' at $z<10$ \citep[see][]{Scannapieco2005}.

Chemical feedback coupled with changing effects of the photoionizing
background on the star formation efficiency in small halos may play an
important role in reconciling low-redshift observations.  Namely, {\it
all} mini-halos must form stars efficiently in order to achieve an
early redshift of reionization \cite[e.g.][]{Cen2005,SBL2003}.  On the
other hand, in order to reconcile the dark halo mass function slope
with the faint luminosity function slope at $z\sim 0$, star formation
must be inhibited in small halos \cite[see, e.g.  ][]{BKW2000}.  This
low-z vs.  high-z conundrum suggests that some transition mechanism
must exist between the nature of star formation at high redshift and
low redshift.  Constraints on the faint end slope of the luminosity
function at high redshifts and observations of how this slope evolves
with time will be important in order to understand what gives rise to
such an effect \cite{Dave2005,Nagamine2005,Bouwens2005c}.

An important driver for the next generation of space and ground-based
telescopes is to directly detect the objects responsible for
reionizing the universe.  As stressed by both Dav\'e and Nagamine, the
existence of massive galaxies at $z\sim 6-8$ is not necessarily a
death blow to LCDM simulations, but may require some substantial
reworking of feedback and star formation prescription if old massive
galaxies at $z\sim 6$ are the norm.  Hydrogen Lyman $\alpha$ is our
primary emission-line probe of the first galaxies.  However, even the
best simulations have difficulties predicting abundances of Lyman
$\alpha$ emitters because of the poorly understood process of Lyman
$\alpha$ escape in these systems
\cite{Barton2004,Dave2005,Nagamine2005}.  Mori \& Umemura argue that
Lyman alpha emitters may represent cooling shocks from high-density SN
driven winds and may represent the future sites of LBGs
\cite{Mori2005}.  But robust predictions for the total Lyman-$\alpha$
emitter abundances remain elusive for the reasons just mentioned.

Oh's talk focused on the issue of escape fraction of Lyman-$\alpha$
photons, and described his work with Hansen on radiative transfer
within (perhaps) the most realistic medium - a multi-phase medium of
dusty, moving optically thick gas clumps.  They show that
Lyman-$\alpha$ can escape from multi-phase dusty galaxies for HI
column densities that would be suppressed for a single-phase medium
\cite{Oh2005}.  They provide a fast method for dealing with this
problem which would seem to be quite useful for simulations of the
type discussed by Dav\'e and Nagamine among others.

A logical place to directly look for the effects of the high-z and
low-Z processes responsible for ionization is in the most metal free
stars in our vicinity: metal poor stars in the stellar halo.  Jason
Tumlinson \cite{Tumlinson2005} has constructed a monte-carlo model of
Milky Way formation in order to test how the low-Z tail of halo stars
may be used as a probe of the $Z_{cr}$ transition and the nature of
the PopIII IMF.  He concludes that the halo star metallicity
distribution function (MDF) is best fit by a PopIII IMF that is not
filled with ultra massive stars.  Rather, he favors a Pop III IMF
devoid of stars smaller than one solar mass, but with a more standard
slope above that point.  Several questions are raised by the
stimulating work of Tumlinson: do we expect the progenitors of the
first metal poor stars stars in our vicinity, or should these stars be
buried in the center of the bulge?  What would be the payoff of mining
the MDF of nearby dwarf galaxies, which likely collapsed quite early?

\section{The Future of First Light and Reionization}

A wide array of new studies and new projects are currently being
planned for the study of first light.

\subsection{The CMB}

Ned Wright pointed out that the WMAP data will get better in the
future -- the satellite is slated for 6-8 years of operation. The
large angles are noise dominated and the error bars will shrink with
time. Wright also pointed out that Planck may have a difficult time
detecting the large angle signal because of 1/f noise.

Keating talked about a near-term experiment called BICEP
\footnote{$www.astro.caltech.edu/\sim lgg/bicep\_front.htm$} that will
map out the polarization anisotropy of the CMB with an angular
precision of about a degree (at 100 and 150 GHz) over about 3\% of the
sky. BICEP will be sensitive to both the E-mode signal from last
scattering and reionization, as well as the B-mode tensor signature on
degree scales and larger.  Another experiment that will produce
results in the near future is QUAD, which concentrates on angular
scales of a degree or smaller.

The B-mode signal and reionization are intimately linked. Reionization
creates a B-mode ``bump'' at large angles analogous to the TE (and EE)
reionization signal at large angles. Measuring this signal can
significantly boost detection sensitivity -- the minimum measurable
energy scale of inflation scales scales as $\tau^{1/4}$
\cite{kaplinghat03} (optimistic scenario neglecting
foregrounds). Measuring these small correlation signals at large
angular separations is a very tough proposition, but the potential
pay-offs are enormous. A positive B-mode measurement would imply
that we have measured the inflationary energy scale.

\subsection{Lyman break and Lyman $\alpha$ searches}

Many future instruments and facilities will study
high-redshift LBGs.  McCarthy described a future Magellan instrument,
4STAR, that will survey an 11$^{\prime} \times 11^{\prime}$ portion of
sky from 0.9-2.5 $\mu$m.  Jensen described an active future at Gemini,
including near-infrared imaging with wide-field multi-conjugate
adaptive optics feeding both FLAMINGOS-2 and GSAOI.

Finding very high redshift galaxies through the use of photometric
redshifts is one key technique of the planned {\it James Webb Space
Telescope} (JWST).  Now set to launch in 2013, Marcia Rieke described
the instruments that JWST will carry.  Specifically designed to detect
first-light objects, NIRcam, is a $0.6-5$ $\mu$m camera designed with
simultaneous long- and short- wavelength channels.  It will fly with 7
broad/medium band filters specifically designed for accurate
photometric redshifts at high $z$.  With $3-5$ $\mu$m sensitivity 100
times better than Spitzer, both Rieke and Rogier Windhorst explained
JWST may detect single young star clusters and superstar clusters; it
will see large PopIII star clusters and dwarf galaxies in the
continuum up to $z=15-25$ \cite{Windhorst2005}.  As Tom Abel
indicated, Population III supernovae may be an abundant populaton
suitable for detection with JWST.  The current aperture of JWST
must be preserved to retain these capabilities.

For the ground-based perspective, McCarthy noted that the mirror
casting has begun for the Giant Magellan Telescope, expecting first
light in the vicinity of 2013.  Similarly, the TMT project, a
collaboration of Caltech, the University of California, NOAO, and
Canada, plans first light in 2014.

For narrow-band searches in the
near-term, Joe Jensen described plans to outfit Gemini with additional
instrumentation and capabilities that will allow the detection of
high-redshift Lyman $\alpha$.  In particular, the observatory plans to
build narrow background-reducing filters for the FLAMINGOS-2
spectrograph and GSAOI.  In addition, P.I. Roberto Abraham is
developing a tunable filter for FLAMINGOS-2 that will search for
lensed $z \gtrsim 7$ Lyman $\alpha$ sources around massive galaxy
clusters (Jensen 2005).

The VLT is also building new instrumentation for the study of first
light through Lyman $\alpha$ searches.  Joss Bland-Hawthorn described
the Dark Ages Lyman $\alpha$ Explorer (DAZLE), a near-infrared
narrow-band imager that exploits new technology for building $R=1000$
filters.  By careful filter-matching and switching between filters,
the DAZLE teams plans to achieve a sensitivity of $2 \times 10^{-18}$
erg s$^{-1}$ cm$^{-2}$ over a large ($6.^{\prime}9$) field of view.
They expect first light with DAZLE in January 2006 (see Bland-Hawthorn
2005).  Bland-Hawthorn also described potential new technologies for
OH suppression and narrow-band imaging over wider fields of view in
the near infrared, including fiber Bragg grating technology and a
beam-correcting narrow-band filter design that can work over a degree
field of view \cite{Hawthorn2005}.  With strong OH suppression and
adaptive optics, Bland-Hawthorn pointed out that 8-meter class
telescopes have the potential to be competitive with JWST.

JWST will contribute to emission-line searches for first light by both
finding the sources themselves with the $R \sim 100$ tunable filter
planned for the Fine Guidance Sensor described by Rieke and by
providing photometric redshifts for the sources and for the
surrounding large-scale structure.

As McCarthy pointed out for the GMT, the planned large ground-based
telescopes, including both the GMT and the TMT, stand to make strong
contributions to probing for early emission-line objects.  These
facilities may be capable of moderate-field Lyman $\alpha$ searches
and/or more targeted searches around dense clusters of galaxies where
the intergalactic medium (IGM) is likely ionized
\cite[e.g.,][]{Furlanetto2005,furl-rec}.  In addition, because the highest
resolution that JWST will provide is $R \sim 3000$, large ground-based
facilities will contribute high-resolution spectroscopy to probe the
intergalactic medium surrounding high-$z$ sources using the profile of
Lyman $\alpha$ line \cite[e.g., ][]{Haiman2002}, and to search for HeI
emission ($\lambda$1640~\AA) that signifies massive star formation
\cite[e.g.,][]{Bromm2001}.

\subsection{Galaxies in the sub-mm}

 At present, Spitzer detection of $z \sim 6$ galaxies and the CO
detection of a $z=6.42$ quasar host galaxy \cite{Walter2004} hold the
records for the detection long-wavelength flux from highest redshift
galaxies.  Because as much as half of the energy from stars and black
holes over the history of the universe is detectable in the
mid-to-far-infrared regime, one focus of future space missions is more
constraints in the far-infrared and sub-mm wavelengths.  Matt Bradford
described an instrument, ``Background-Limited Infrared-Submillimeter
Spectrograph'', or {\it BLISS}, for the planned Japanese SPICA mission
set to launch in 2013 \cite{Bradford2005}.  With a 3.5-m mirror and
instruments actively cooled to 4.5K at L2, SPICA will offer huge gains
in sensitivity over existing facility in the 40-600 $\mu$m.  The
mission will bring far-IR emission lines into the range of
detectability for $z=5$ galaxies with ULIRG-type spectra, allowing
diagnostics for star formation in molecular gas, black holes, gas
density, the hardness of the stellar radiation field in distant
galaxies(hence stellar mass and starburst age), and elemental
abundances.  Looking even further into the future, Martin Harwit
described a far-IR/submm km-baseline interferometer in space, the
``Submillimeter Probe of the Evolution of Cosmic Structure'', or {\it
SPECS}, a space-based interferometer array with two 4-meter elements
and variable baselines extending as far as 1 km \cite{Harwit2005}.
One key science driver for the mission is to see and resolve silicate
features from individual pair-instability supernovae as high as $z \sim 20$,
at 200 $\mu$m.  The project requires technology development, including
tethered flight of interferometer, and development of low-noise
detector technology

\subsection{X-rays}

With an effective area of 100 m$^2$ and an angular
resolution of $0.^{\prime\prime}1$, Gen-X is a proposal for the
next-generation x-ray satellite, as described by Rogier Windhorst.
The science goals include spectra of x-ray afterglows of gamma-ray
bursts and early supernovae, early black holes at $z \sim 15$,
intermediate-mass black holes at $1 \lesssim z \lesssim 3$, embedded
star formation rates in $z \sim 3$ galaxies, and supermassive black
holes in the centers of galaxies to $z \sim 10$
\cite{Windhorst2005GenX}.

\subsection{Gravitational Waves}

If there is a significant population of merging black holes at
redshifts around the era of reionization, Naoki Seto described how
gravitational wave observations with the Laser Interferometer Space
Antenna (LISA) may be able to extract parameters of these binaries. If
the high redshift sample is large enough, it may also be possible to
perform statistical studies to distinguish various models of the
initial mass function of black holes and their binaries 
\cite{Koushiappas2005}.
	
\subsection{Fluctuations in the far-IR background} 

Deep IR imaging between 1 $\mu$m and 2 $\mu$m is clearly needed to
establish the contribution from Lyman-$\alpha$ galaxies. While such
imaging will eventually be realized with instruments on JWST, the
approach of Kashlinsky et al. \cite{Kashlinsky05} can be implemented
at lower wavelengths with an IR imaging camera on a rocket-borne
experiment. While ground-based observations can also be considered,
such observations are always affected by varying atmospheric noise,
especially at tens of arcminute scales or more where signatures
related to high-redshift anisotropy fluctuations are expected.  The
experiment discussed in \cite{Bock05}, CIBER: Cosmic Infrared
Background Experiment, attempts to image the IR sky above 1 $\mu$m
over a several square degrees down to a point source detection limit
around 19th magnitude.  CIBER consists of a wide-field two-color
camera, a low-resolution absolute spectrometer, and a high-resolution
narrow-band imaging spectrometer.

In a short rocket flight CIBER will have the sensitivity to probe fluctuations
$100 \times$ fainter than IRTS/DIRBE with sufficient resolution to
remove local-galaxy correlations.  The low-resolution spectrometer is
designed to search for a redshifted Lyman cutoff feature between 0.8 -
2.0 $\mu m$, while the high-resolution spectrometer will trace
zodiacal light using the intensity of scattered Fraunhofer lines,
providing an independent measurement of the zodiacal emission and a
new check of DIRBE zodiacal dust models \cite{Kelsall98}.
Current zodiacal model discrepancies based on DIRBE are at the $\sim
10 \%$ level, producing large discrepancies in the derived IRB. CIBER
could provide the decisive data to clearly establish the exact amount
of excess IRB, and when combined with wide-field imaging data as well
as data from Spitzer, the surface density of galaxies. Having an early
understanding of the expected population of galaxies at redshifts
greater than 8 is crucial to the successful implementation of
observations with near-IR instruments on JWST.

\subsection{21 cm measurements}

The 21 cm line of neutral Hydrogen provides one of the most powerful
probes of reionization history. When compared to CMB anisotropies that
provide an integrated measure of the reionization history, in the form
of an optical depth to electron scattering, the spin-flip transition
of neutral Hydrogen at a rest wavelength of 21 cm may allow, at least
in principle, this history to be established as a function of 
redshift.  The scientific motivation to study 21 cm fluctuations are
described in \cite{Furlanetto05} and \cite{Carilli05}.  The 21 cm
anisotropies probe fluctuations in the neutral Hydrogen density and
gas temperature fields during the era of reionization. The models
related to these underlying fluctuations are now studied both with
analytical techniques as well as numerical simulations of
reionization. In addition to the neutral density field associated with
the intergalactic medium before complete reionization, clustered
regions of neutral gas, in the form of minihalos \cite{Ahn05}, provide
unique signatures in 21 cm anisotropy maps. It is expected that under
certain conditions this minihalo contribution may be the dominant
signal in 21 cm fluctuations.

In addition to probing physics during reionization through
anisotropies, 21 cm fluctuations also provide several unique
applications in cosmology.  The small scale anisotropies probed by the
density field are sensitive to details of the small scale dark matter
fluctuations and could be used to probe more accurately cosmological
parameters such as the neutrino mass or primordial non-Gaussianity
\cite{Loeb03}. The intervening large-scale structure,
between us and the reionization surface, gravitationally lenses and
modifies the anisotropy pattern. A study of such lensed anisotropies
can be used to map integrated dark matter fluctuations out to a
redshift higher than possible with galaxy imaging data. The same
lensing measurements could be used to {\it clean} CMB polarization
maps and account for the lensing B-modes that can confuse the
detection of a primordial gravitational wave background \cite{Sigurdson05}.

While 21 cm studies have numerous applications, in practice, there are
several difficulties to overcome. These involve both experimental
challenges related to achieving high sensitivity observations at very
low frequencies, where man-made interference must be overcome, as well
separating the high redshift signal from highly contaminating
foreground radio emission. Such foregrounds include synchrotron and
free-free emission from the Milky Way and low-frequency radio point
sources and free-free emission from free electrons in the IGM. The
frequency dependent variations in the anisotropy structure of the 21
cm emission, while the foregrounds remain essentially the same, may
allow a separation between the two \cite{santos05}. Practical issues
related to 21 cm observations, especially with the planned Mileura
Wide-field Array (MWA), are discussed by Miguel Morales
\cite{Morales05}, while various other observational programs were
discussed by Jeff Peterson and Chris Carilli \cite{Carilli05}.

\section{Summary}

As Virginia Trimble describes, the quest for an understanding of the
first light in the universe has long been a subject of interest
\cite{Trimble05}.  Progress to date is rapidly accelerating as the
CMB, the infrared background light, and high-redshift galaxies are
already yielding valuable clues about the process of reionization.
The strong, well-motivated theoretical understanding of these
processes that is currently being developed is vital to synthesizing a
coherent picture for reionization and early structure formation.
However, because of uncertainties in the existing models, theory alone
is not enough to drive the field forward.  Fortunately, the vast array
of exciting new studies and facilities being planned will yield
unprecedented information about the time in the early phases of cosmic
history when light was first generated from stars in the universe.

\bigskip

We thank Casey Papovich and Aaron Barth for insightful comments.  This
conference and summary were supported by the University of California,
Irvine.  In particular, we gratefully acknowledge generous donations
from the Office of Research and Graduate Studies and the Office of the
Dean of Physical Sciences.

\newcommand{\apjl}{{\it Astrophy. J. Lett.}}
\newcommand{\apj}{{\it Astrophy. J.}}
\newcommand{\aj}{{\it Astronom. J.}}
\newcommand{\mnras}{{\it MNRAS}}


\begin{thebibliography}{99}
\frenchspacing

\bibitem{Abraham2004}
R. G. Abraham {\it et al.}, \aj, {\bf 127}, 2455 (2004).

\bibitem{Ahn05}
K. Ahn {\it et al.}, these proceedings (2006).

\bibitem{Barton2004}
E.J. Barton, R. Dav\'{e}, J.-D. T. Smith, C. Papovich, L. Hernquist,
\& V. Springel, \apjl, {\bf 604}, L1 (2004).

\bibitem{Becker2001}
R. H. Becker, {\it et al.}, \aj, {\bf 122}, 2850 (2001).

\bibitem{Hawthorn2005}
J. Bland-Hawthorn, these proceedings (2006).

\bibitem{Bock05}
J. J. Bock {\it et al.}, these proceedings (2006).

\bibitem{bond84}
J.~R. Bond, \& G. Efstathiou, 1984, \apjl, {\bf 285}, L45

\bibitem{Bouwens2004}
R. J. Bouwens, {\it et al.}, \apj, {\bf 616}, 79 (2004).

\bibitem{Bouwens2005b}
R. Bouwens \& G. Illingworth, these proceedings (2006).

\bibitem{Bouwens2005a}
R. J. Bouwens, G. D. Illingworth, J. P. Blakeslee, \& M. Franx, \apj,
{\bf 624}, 5 (2005a).

\bibitem{Bouwens2005c}
R. J. Bouwens, G. D. Illingworth, J. P. Blakeslee, \& M. Franx, 
\apj, in press, astro-ph/0509641 (2005b).

\bibitem{Bradford2005}
C. M. Bradford, T. Nakagawa, \& the BLISS-SPICA
team, these proceedings (2006).

\bibitem{Bremer2004}
M. N. Bremer, J. B. Jensen, M. D. Lehnert, N. M. F\"{o}rster
Schreiber, \& L. Douglas, \apj, {\bf 615}, 1 (2004).

\bibitem{Bromm2001}
V. Bromm, R. P. Kudritzki, \& A. Loeb, \apj, {\bf 552} ,464 (2001).

\bibitem{BKW2000} Bullock, J.~S.,
Kravtsov, A.~V., \& Weinberg, D.~H., \apj, {\bf 539}, 517 (2000).

\bibitem{Bunker2005}
A. Bunker, E. Stanway, R. Ellis, R. McMahon, L. Eyles, \& M. 
Lacey, these proceedings (2006).

\bibitem{Bunker2004}
A. Bunker, E. R. Stanway, R. S. Ellis, \& R. G. McMahon, \mnras,
{\bf 355}, 374 (2004).

\bibitem{Carilli05}
C. Carilli, these proceedings (2006).

\bibitem{Cen2005}
R. Cen, these proceedings (2006).

\bibitem{Cooke2005}
J. Cooke, A. M. Wolfe, J. X. Prochaska, \& 
E. Gawiser, \apj, {\bf 621}, 596 (2005).

\bibitem{Cooke2005b}
J. Cooke, A. M. Wolfe, J. X. Prochaska, \& E. Gawiser, 
these proceedings (2006).

\bibitem{Cooray04a}
A.~Cooray {\it et al.}, \apj, {\bf 606}, 611 (2004).

\bibitem{Cooray04b}
A.~Cooray \& N.~Yoshida, \mnras, {\bf 51}, L71 (2004).

\bibitem{Cooray05}
A.~Cooray, J. Bock, I. Sullivan, \& B. Keating, in preparation (2006).

\bibitem{Dave2005}
R. Dav\'e, these proceedings (2006).

\bibitem{Dijkstra2005}
M. Dijkstra, these proceedings (2006).

\bibitem{Djorgovski2005}
S. G. Djorgovski, M. Bogosavljevic, \& A. Mahabal, these proceedings (2006).

\bibitem{Fan2001}
X. Fan, {\it et al.}, \aj, {\bf 122}, 2833 (2001).

\bibitem{Fan2002}
X. Fan, {\it et al.}, \aj, {\bf 123}, 1247 (2002).

\bibitem{Furlanetto2005}
S. R. Furlanetto, M. Zaldarriaga, \& L. Hernquist, \mnras, in press (2006).

\bibitem{furl-rec}
S.~R. Furlanetto \& S.~P. Oh, \mnras, {\bf 363}, 1031 (2005).

\bibitem{furlanetto04} 
S. R. Furlanetto, M. Zaldarriaga, \& L. Hernquist, \apj, {\bf 613}, 1 (2004).

\bibitem{Furlanetto05}
S. R. Furlanetto, these proceedings (2006).

\bibitem{Giavalisco2004}
M. Giavalisco, {\it et al.}, \apj, {\bf 600}, 103 (2004).

\bibitem{Haiman2002}
Z. Haiman, \apj Lett., {\bf 576}, 1 (2002).

\bibitem{Oh2005}
M. Hansen \& S. Peng Oh, these proceedings (2006).

\bibitem{Harwit2005}
M. Harwit, D. Leisawitz, \& S. Rinehart, these proceedings (2006).

\bibitem{Hauser98}
M.~G.~Hauser {\it et al.}, \apj, {\bf 508}, 25 (1998).

\bibitem{Hauser01}
M.~G.~Hauser \& E.~Dwek, ARAA {\bf 39}, 249 (2001).

\bibitem{Jensen2005}
J. B. Jensen, these proceedings (2006).

\bibitem{kaplinghat03}
M.~Kaplinghat, L.~Knox, \& Y.~S.~Song, {\it Phys.\
  Rev.\ Lett.}, {\bf 91}, 1301 (2003)

\bibitem{kaplinghat02}
M.~Kaplinghat, M.~Chu, Z.~Haiman, G.~Holder,
  L.~Knox, \& C.~Skordis, Astrophys.\ J.\  {\bf 583}, 24 (2003)

\bibitem{Kashlinsky04}
A.~Kashlinsky {\it et al.}, \apj, {\bf 608}, 1 (2004).

\bibitem{Kashlinsky05}
A.~Kashlinsky {\it et al.}, these proceedings (2006).

\bibitem{Keating2005}
B. Keating \& N. Miller, these proceedings (2006).

\bibitem{Kelsall98}
T.~Kelsall {\it et al.}, \apj, {\bf 508}, 44 (1998).

\bibitem{Kogut2003}
A. Kogut, {\it et al.}, \apj, Supp. {\bf 148}, 161 (2003).

\bibitem{Koushiappas2005}
S. M. Koushiappas \& A. R. Zentner, \apj, in press, astro-ph/0503511 (2005).

\bibitem{Loeb03}
A.~Loeb, M.~Zaldarriaga, {\it Phys. Rev. Lett.}, {\bf 92}, 211301 (2004).

\bibitem{Madau05}
P.~Madau \& J.~Silk, \mnras, {\bf 359}, 37 (2005).

\bibitem{Malhotra2005}
S. Malhotra, {\it et al.}, \apj, {\bf 626}, 666 (2005).

\bibitem{Martin2005}
C. L. Martin, M. Sawicki, A. Dressler, \& P. J. McCarthy, these 
proceedings (2006).

\bibitem{Matsumoto05}
T.~Matsumoto {\it et al.}, \apj, {\bf 626}, 31 (2005).

\bibitem{McCarthy2004}
P. J. McCarthy {\it et al.}, \apj, {\bf 614}, 9 (2004).

\bibitem{mcquinn05a} 
M. McQuinn, S.~R. Furlanetto, L. Hernquist,
  O. Zahn, \& M. Zaldarriaga, \apj, 630, 643 (2005).

\bibitem{mcquinn05b} 
M. McQuinn, S.~R. Furlanetto, L. Hernquist,
  O. Zahn, \& M. Zaldarriaga, M., these proceedings (2006).

\bibitem{Mesinger2005}
A. Mesinger, these proceedings (2006).

\bibitem{Mobasher2005}
B. Mobasher, {\it et al.}, \apj, in press, astro-ph/0509768 (2006).

\bibitem{Morales05}
M. F. Morales {\it et al.}, these proceedings (2006).

\bibitem{Mori2005}
M. Mori \& M. Umemura, these proceedings (2006).

\bibitem{Nagamine2005}
K. Nagamine, these proceedings (2006).

\bibitem{Panagia2005}
N. Panagia, {\it et al.}, \apjl, {\bf 633}, L1 (2005).

\bibitem{Papovich2005}
C. Papovich and the GOODS and MIPS GTO teams, these proceedings (2006).

\bibitem{Pello2004}
R. Pell\'{o}, D. Schaerer, J. Richard, J.-F. le Borgne, \& J.-P. Kneib,
{\it A\&A}, {\bf 416}, 35 (2004).

\bibitem{polnarev85} 
A. G. Polnarev, {\it Soviet Astronomy}, {\bf 29}, 607 (1985).

\bibitem{Rhoads2004}
J. E. Rhoads, {\it et al.}, \apj, {\bf 611}, 59  (2004).

\bibitem{Santos2004}
M. R. Santos, \mnras, {\bf 349}, 1137 (2004).

\bibitem{santos03} 
M. G. Santos, A. Cooray, Z. Haiman, L. Knox, \&
  C. -P. Ma, \apj, {\bf 598}, 756 (2003).

\bibitem{santos05}
M. G. Santos,  A. Cooray, L. Knox, \apj, {\bf 625}, 575 (2005).

\bibitem{Scannapieco2005}
E. Scannapieco, these proceedings (2006).

\bibitem{Schneider2005}
R. Schneider, these proceedings (2006).

\bibitem{Sigurdson05}
K.~Sigurdson, \& A.~Cooray, submitted, astro-ph/0502549 (2006).

\bibitem{SBL2003} R. S. Somerville,
J. S. Bullock, \& M. Livio, \apj, {\bf 593}, 616 (2003).

\bibitem{Stanway2004}
E. R. Stanway {\it et al.}, \apjl, {\bf 604}, L13, (2004).

\bibitem{Stark2005}
D. P. Stark, \& R. S. Ellis, these proceedings (2006).

\bibitem{Steidel93}
C. C. Steidel, \& D. Hamilton, \aj, {\bf 105}, 2017 (1993).

\bibitem{Tinsley73}
B. Tinsley, {\it Astrophys. Lett.}, {\bf 14}, 15 (1973).

\bibitem{Trimble05}
V. Trimble, these proceedings (2006).

\bibitem{Tumlinson2005}
J. Tumlinson, these proceedings (2006).

\bibitem{Venkatesan2005}
A. Venkatesan, these proceedings (2006).

\bibitem{Walter2004}
F. Walter, {\it et al.} \apj, {\bf 615}, 17 (2004).

\bibitem{Willis2005b}
J. P. Willis, F. Courbin, J.-P. Kneib, \& D. Minniti, these proceedings (2006).  

\bibitem{Willis2005a}
J. P. Willis, \& F. Courbin, \mnras, {\bf 357}, 1348 (2005).  

\bibitem{Windhorst2005}
R. A. Windhorst, S. H. Cohen, R. A. Jansen, 
C. Conselice, \& H. Yan, these proceedings (2006).

\bibitem{Windhorst2005GenX}
R. A. Windhorst {\it et al.}, these proceedings (2006).

\bibitem{Xu02}
J.~Xu {\it et al.}, \apj, {\bf 580}, 653 (2002).

\bibitem{Yan2005a}
H. Yan, {\it et al.} \apj, {\bf 636}, 109 (2005).

\bibitem{Yan2005b}
H. Yan, {\it et al.}, these proceedings (2006).

\bibitem{Yoshida2005}
N. Yoshida, these proceedings (2006).

\bibitem{zaldarriaga95}
M.~Zaldarriaga \& D.~D.~Harari, {\it Phys.\ Rev.\ D},
  {\bf 52}, 3276 (1995)

\bibitem{zaldarriaga96}
M.~Zaldarriaga, Phys.\ Rev.\ D {\bf 55}, 1822 (1997)

\bibitem{zaldarriaga97}
M.~Zaldarriaga, D.~N.~Spergel, \& U.~Seljak,  Astrophys.\ J.\  
{\bf 488}, 1 (1997)



\end{thebibliography}
\end{document}